\documentclass[prl,twocolumn,amsfonts,superscriptaddress,showpacs]{revtex4}
 
\usepackage[]{graphicx} 
\usepackage{amsbsy} 
\usepackage{float}
\usepackage{color}
\usepackage{subfigure}
 
\newcommand{\bea}{\begin{eqnarray}}
\newcommand{\eea}{\end{eqnarray}}
\newcommand{\ba}{\begin{eqnarray*}}
\newcommand{\ea}{\end{eqnarray*}}

\begin{document} 
\title{Surface effects on the Mott-Hubbard transition in archetypal V$_2$O$_3$}

\author{G.~Lantz} 
\affiliation{Laboratoire de Physique des Solides, CNRS-UMR 8502, Universit\'{e} Paris-Sud, F-91405 Orsay, France}
\author{M.~Hajlaoui} 
\affiliation{Laboratoire de Physique des Solides, CNRS-UMR 8502, Universit\'{e} Paris-Sud, F-91405 Orsay, France}
\author{E.~Papalazarou} 
\affiliation{Laboratoire de Physique des Solides, CNRS-UMR 8502, Universit\'{e} Paris-Sud, F-91405 Orsay, France}
\author{V.~L.~R.~Jacques}
\affiliation{Laboratoire de Physique des Solides, CNRS-UMR 8502, Universit\'{e} Paris-Sud, F-91405 Orsay, France}
\author{A.~Mazzotti}
\affiliation{Laboratoire de Physique des Solides, CNRS-UMR 8502, Universit\'{e} Paris-Sud, F-91405 Orsay, France}
\author{M.~Marsi}
\affiliation{Laboratoire de Physique des Solides, CNRS-UMR 8502, Universit\'{e} Paris-Sud, F-91405 Orsay, France}
\author{S.~Lupi}
\affiliation{Dipartimento di Fisica, Universit\`{a} di Roma La Sapienza, Piazzale A.~Moro, I-00185 Roma, Italy }
\author{M.~Amati}
\affiliation{Elettra-Sincrotrone Trieste S.C.p.A., SS14 - km 163.5 in AREA Science Park, 34149 Basovizza, Trieste, Italy}
\author{L.~Gregoratti}
\affiliation{Elettra-Sincrotrone Trieste S.C.p.A., SS14 - km 163.5 in AREA Science Park, 34149 Basovizza, Trieste, Italy}
\author{L.~Si} 
\affiliation{Institute for Solid State Physics, Vienna University of Technology, AT-1040 Vienna, Austria}
\author{Z.~Zhong}
\affiliation{Institute for Solid State Physics, Vienna University of Technology, AT-1040 Vienna, Austria}
\affiliation{Institute for Theoretical and Astrophysics, University of W\"urzburg, Am Hubland 9704, W\"urzburg, Germany}
\author{K.~Held}
\affiliation{Institute for Solid State Physics, Vienna University of Technology, AT-1040 Vienna, Austria}
\date{\today} 
 
\begin{abstract} 
We present an experimental and theoretical study exploring surface effects on the evolution of the metal-insulator transition in the model Mott-Hubbard compound Cr-doped V$_2$O$_3$. We find a microscopic domain formation that is clearly affected by the surface crystallographic orientation. Using scanning photoelectron microscopy and X-ray diffraction, we find that surface defects act as nucleation centers for the formation of domains at the temperature-induced isostructural transition and favor the formation of microscopic metallic regions. A density functional theory plus dynamical mean field theory study of different surface terminations shows that the surface reconstruction with excess vanadyl cations leads to doped, and hence more metallic surface states, explaining our experimental observations.
\end{abstract} 
 
\maketitle 
Metal to insulator transitions (MIT) are among the most remarkable macroscopic effects of electronic correlations in condensed matter. After many experimental and theoretical studies, it has been possible to understand the crucial role played by the lattice, which can stabilize electronic instabilities and guide the evolution of correlation-driven phenomena such as Mott-Hubbard transitions \cite{Imada1998}. However these studies have concentrated on bulk properties: the surface behavior is rarely discussed and is indeed a more complicated problem. At the surface, the atomic coordination number and screening change and this has an effect on electronic correlations, but other factors such as surface reconstruction and lattice defects can also affect the MIT.

Cr-doped V$_2$O$_3$ is the prototype Mott-Hubbard material \cite{McWhan1969,McWhan1970,Hansmann2013}, presenting a correlation-induced MIT without symmetry breaking. Strong electronic correlation split the noninteracting bands into interacting upper and lower Hubbard bands. In the metallic phase a strongly renormalized quasiparticle peak remains at the Fermi level which is reminiscent of the uncorrelated band structure. The phase diagram of V$_2$O$_3$ consists of three phases: paramagnetic insulator (PI), paramagnetic metal (PM), and antiferromagnetic insulator (AFI) \cite{McWhan1969} (see Fig. \ 1 a)). The Mott transition takes place between the PI and PM phases, and it can be induced by increasing pressure starting from the PI phase, as well as by decreasing temperature for doping levels around 1.1$\%$ Cr concentration. The orbital degrees of freedom in V$_2$O$_3$  have to be taken into account in order to understand the MIT: the low-lying orbitals are the t$_{2g}$ orbitals made up of singly degenerate a$_{1g}$ and doubly degenerate e$^{π}_g$ orbitals in the crystal field of the corundum lattice structure. At the MIT the a$_{1g}$ bands are shifted up in energy and the e$^{π}_g$ bands split into two Hubbard bands, with the Hund’s exchange leading to a local spin alignment throughout the transition \cite{Held2001B,Keller2004,Poteryaev2007,Park2000}. Although the surface of V$_2$O$_3$ has been well studied for the pure compound \cite{Window2011,Feiten2015}, and extensive experimental \cite{Surnev2003} and theoretical investigations \cite{Kolczewski2007} have been carried out on the surface termination of V$_2$O$_3$, very little is known about the effects of the surface on the Mott transition. In this letter, we explore specific surface effects on the evolution of the Mott-transition in (V$_{1-x}$Cr$_x$)$_2$O$_3$ both experimentally (using X-ray diffraction (XRD) and scanning photoelectron microscopy (SPEM)) and theoretically by density functional theory plus dynamical mean field theory (DFT+DMFT) \cite{LDADMFT,Kotliar06,heldreview}

\begin{figure}
\includegraphics[angle=0,width=1\linewidth,clip=true]{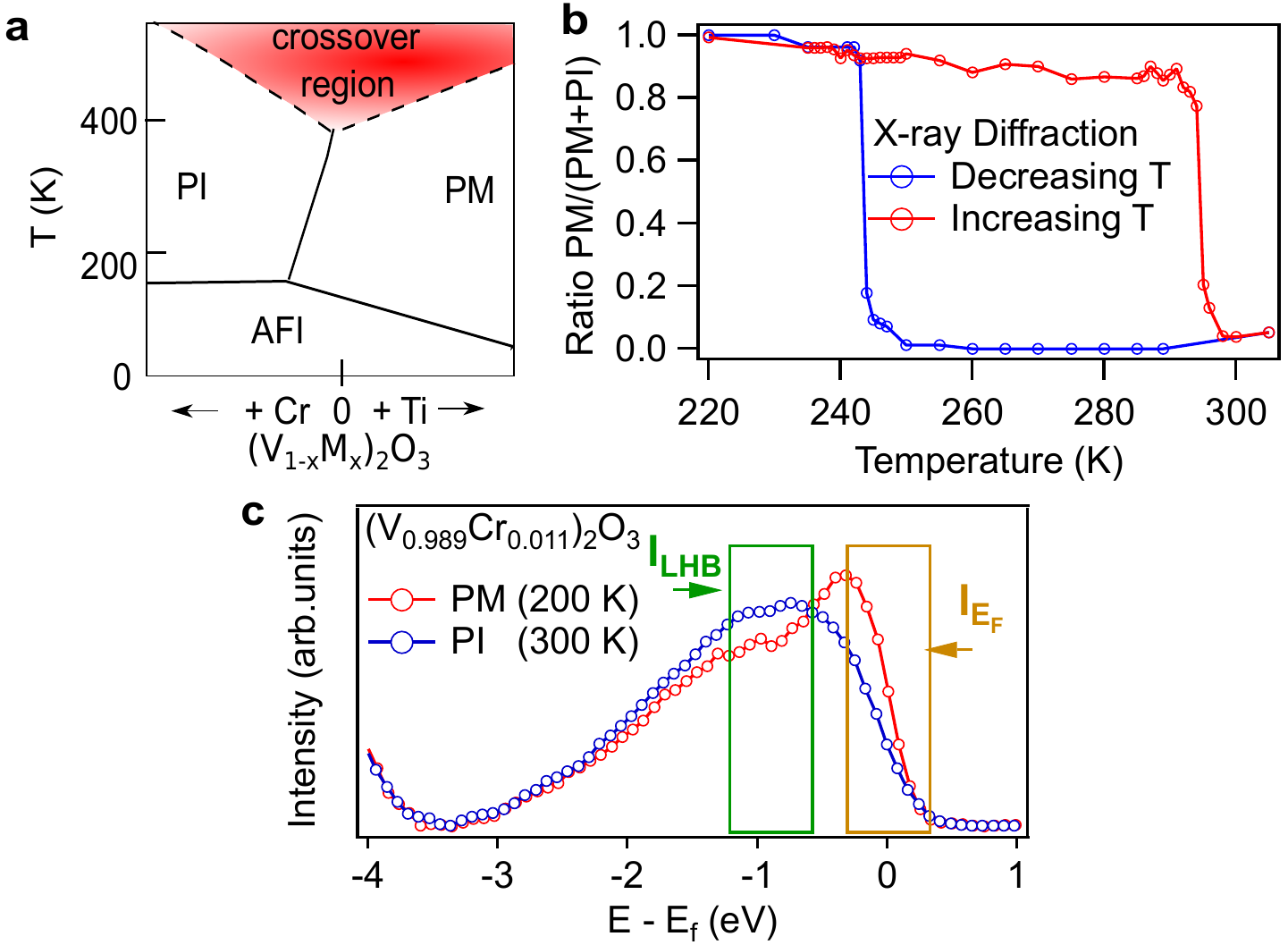}
\caption{\textbf{a)} Phase diagram for (V$_{1-x}$Cr$_x$)$_2$O$_3$.  \textbf{b)} Ratio PM/(PM+PI) versus temperature probed by X-ray diffraction on a monocrystal. \textbf{c)} Photoemission spectra for (V$_{1-x}$Cr$_x$)$_2$O$_3$ at x=0.011, 300 K and 200 K. The increase of spectral weight near the Fermi level indicates the phase transition from insulator to metal when decreasing the temperature. I$_{E_F}$ and I$_{LHB}$ represent respectively the intensity near E$_F$ and of the LHB. The ratio between the two provides a measure for the metallicity of the probed surface.}
\label{spectra} 
\end{figure}

The Mott transition is a first order transition therefore domain coexistence can occur and a large hysteresis is possible. Fig \ 1 b) shows XRD measurements on a monocrystal done during the temperature driven phase transition \cite{xrd1}. A large hysteresis is seen comparable to earlier results \cite{McWhan1970}. However the transition happens in less than 2 K at 244 K whereas in powder XRD the transition lasts 60 K. The major difference between a monocrystal and a powder is the surface to bulk ratio. The MIT at the surface might be greatly effected by defects or surface reconstruction. In order to understand this discrepancy we probe the surface using SPEM.

{\em  Method -- Experimental.} SPEM experiments were performed as a function of temperature on the Escamicroscopy beamline at Elettra synchrotron, using photons of 400 eV \cite{spem1}. SPEM uses a direct approach to photoelectron spectromicroscopy which is the use of a focused photon beam, down to 150 nm in diameter, to illuminate the sample. Photoelectrons are collected with a hemispherical electron analyzer and  detected by a 48-channels electron detector. SPEM can operate in two modes: i) XPS spectroscopy from a sub-micron spot; ii) imaging where the sample surface is mapped within a selected kinetic energy. In imaging mode, all 48 channels are recorded for every pixel of the image. In our experiments a 2 eV range near the Fermi level was chosen to observe the MIT, as seen in Fig. \ref{spectra}. 

As it has already been shown in previous SPEM measurements \cite{Lupi2010}, coexisting PM and PI phases in (V$_{1-x}$Cr$_x$)$_2$O$_3$ can be unambiguously distinguished by
the photoemission signal from the outer valence electronic states, but the physical origin remained mysterious. With the present study we are able to solve this puzzle. The improved experimental conditions allow us to show a clear correlation between the shape and position of domains at the metal-insulator transition, and the fact that surface structural defects actually favor the formation of metallic domains. This was possible thanks to: i) the use of higher energy photons, that make it possible to have a clearer spectroscopic contrast \cite{Mo2003}; ii) an improved spatial resolution (150 nm); and iii) a direct comparative study between different surfaces of the same material. In particular, high quality single crystals from Purdue University were either cleaved in the (001) plane or the (102) plane, keeping always the [001] direction pointing towards the electron analyzer which  was shown to give the strongest quasiparticle peak in the photoemission signal\cite{Rodolakis2009} (see Fig.\ 2 top). The samples were cleaved and measured under UHV conditions (2x10$^{-10}$ mbar or better) to avoid surface contamination (the carbon photoemission peak was regularly checked in the course of the experiment). In order to obtain genuine two-dimensional maps of the PM vs PI concentration at the surface, the photoemission images were corrected for topographic effects \cite{Marsi1997}. As a measure for the metallicity of the sample, we take the ratio between the intensity near E$_F$ (I$_{E_F}$) and near the Lower Hubbard Band (LHB) (I$_{LHB}$), see Fig. \ref{spectra}. The images obtained in this way are real-space "metallicity maps" of the sample surface \cite{Lupi2010}.

\begin{figure}
\includegraphics[angle=0,width=1\linewidth,clip=true]{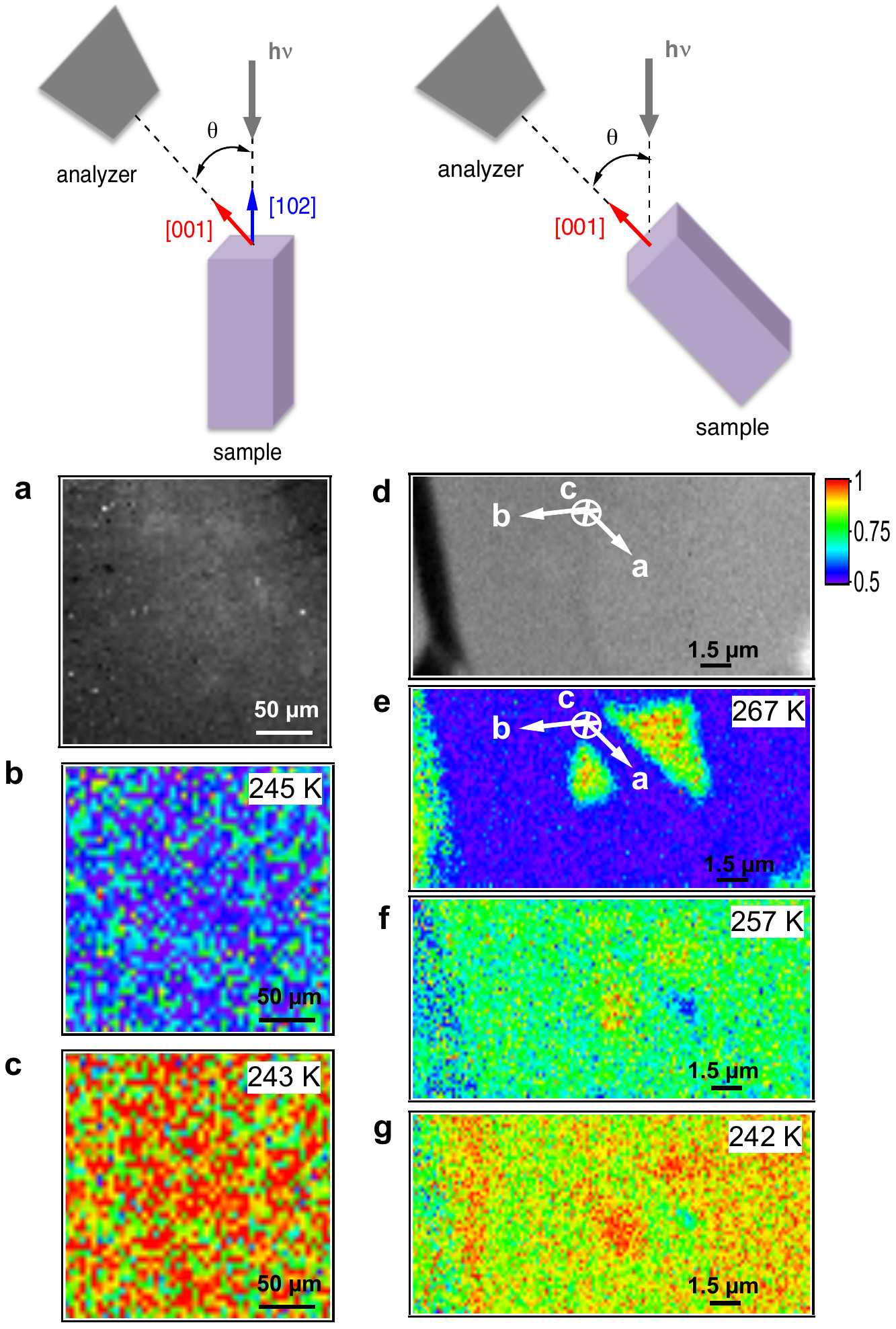}
\caption{\textbf{Top:} Sample orientation versus incoming photons and analyzer \textbf{Left:}  Sample cleaved along the (102) direction with analyzer in (001) direction. \textbf{a)}: Surface morphology image taken at a photoelectron energy corresponding to the V3p core level, and at temperature T = 245 K. \textbf{b)} contrast $I_{E_F}/I_{LHB}$ between the PM and PI phases for 245 K and \textbf{c)} for 243 K. No coexisting domains could be detected during the phase transition. \textbf{Right:}  Sample fractured along the (001) direction  with analyzer in (001) direction.  \textbf{d)}  Surface morphology image. The following images show the contrast between the PM and PI phases: \textbf{e)} T=267 K, \textbf{f)} T=257 K and \textbf{g)} T= 242 K. } 
\label{spem} 
\end{figure}

{\em SPEM results} Fig. \ref{spem} shows a region on the (102) cleaved surface at different temperatures while cooling down the sample. Since the (102) plane is the natural cleaving direction, it can produce large flat surfaces with few defects. In particular, the probed region does not present any significant defects detectable with our spatial resolution (150 nm): this is evident from Fig. \ref{spem} a), where the image has not been corrected for surface morphology and which cannot reveal the presence of any topographic defect at the surface. By observing the metallicity images  we found that no MIT could be seen until 245 K (Fig. \ref{spem} b)), while the whole surface turned metallic at 243 K (Fig. \ref{spem} c)). This indicates that the transition was too sudden for our experimental conditions to detect the presence of domains, the limiting factor being the relatively long acquisition time for each photoemission image and/or the size of the domains. Hence, we can conclude that, within the limits of our spatial resolution, the entire probed region turns metallic in less than 2 K, similar to the bulk probed by our XRD experiment. Therefore the (102) plane behaves like the bulk material.

\begin{figure}
\includegraphics[angle=0,width=1\linewidth,clip=true]{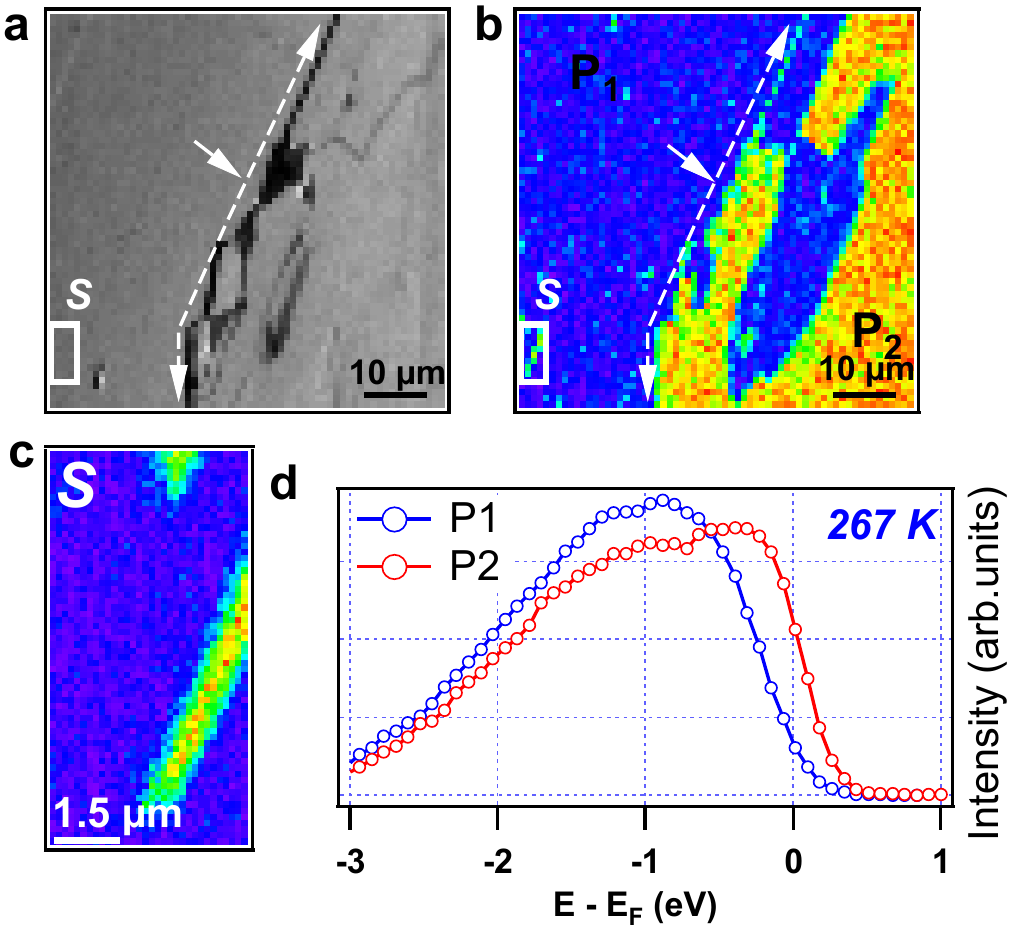}
\caption{Sample cut along the (001) plane at 267 K. {\bf a)} Surface morphology image with a crack marked by the dashed arrows and a smaller defect marked by the $S$ box.  {\bf b)} The corresponding phase contrast images show the apparition of metallic domains starting at 267 K. The large domains are separated by cracks in the sample.  {\bf c)} Close up image of a cleavage step, which presents a more metallic behavior.  {\bf d)} Photoemission spectra from the P$_1$ and P$_2$ points.} 
\label{defects} 
\end{figure}

On the contrary, when cleaving along the (001) direction, we find coexisting domains on the surfaces over a wide temperature range, starting as soon as 267 K. We start by analyzing an area where no structural defects  could be detected with our spatial resolution, as revealed by the uncorrected image in Fig. \ref{spem} d): the first metallic domains appear starting at 267 K (Fig. \ref{spem} e)), and they present a triangular shape, revealing a clear correlation with the hexagonal symmetry of the (001) plane (the a and b crystallographic directions are shown to allow a direct comparison): this strongly suggests that the borders of the metallic domain correspond to surface steps. The rest of the area is still insulating and starts its transformation around 257 K (Fig. \ref{spem} f)). Eventually at 242 K (Fig. \ref{spem} g)), the transition is almost complete, similarly to the (102) plane. On the (001) surface, the sample also has regions with a significant amount of large defects as represented in Fig. \ref{defects}. Here large metallic domains appear between the cracks. Let us now focus our attention on smaller defects, most likely corresponding to cleavage steps, such as the one found in the "S" area shown in detail in Fig. \ref{defects}(c): the metallic domains appear to follow the cleavage step direction. Overall, metallic domains are present for the (001) surface well above 243 K, which is the transition temperature of the (102) plane.

Our experimental findings confirm that structural defects can act as nucleating centers for the insulator-to-metal transition, thus guiding the evolution of the formation of domains during the coexistence\cite{Lupi2010}. Surprisingly, they also clearly indicate that structural defects at the surface of (V$_{1-x}$Cr$_x$)$_2$O$_3$ favor the formation of a metallic phase over an insulating one.

{\em Theory.}
One might envisage the cleavage steps as an extra surface so that surface effects should be enhanced. However the usual suspects of electronic correlations at (enhanced) surfaces, i.e., an enhanced $U$ because of the reduced screening and the removed hopping perpendicular to the surface, suggest the surface and cleavage steps to be more insulating. We find the contrary; they are more metallic.

With the simple explanations at failure, we need to inspect the surface more thoroughly, including the proper surface reconstruction. DFT calculations \cite{Kresse} indicate, depending on the oxygen partial pressure, a VO surface termination with various excess of vanadyl or a O$_3$ termination. The former has also been identified in some experiments \cite{Surnev2003,Kolczewski2007,Dupuis2003,Schoiswohl2004,Feiten2015}. Fig.\ 4 shows the actual supercells with  VO and  O$_3$ termination that we consider in our calcualtion. One can envisage the surface as consisting of several layers of the stoichiometric (V-O$_3$-V) unit cell, i.e.,  (V-O$_3$-V) (V-O$_3$-V) $\cdots$ (V-O$_3$-V) plus an extra  V or (O$_3$-V) layer for the
 VO and O$_3$ termination, respectively, see Fig.\ 4 (upper panel).

{\em Method -- Theory.} We perform a full relaxation of the VO and O$_3$ terminated V$_2$O$_3$ surfaces in the corundum structure including a 12 {\AA } thick vacuum layer, using the Vienna ab-initio simulation package (VASP) \cite{7,8}, with GGA-PBE functional \cite{9}. In the case of the   O$_3$ termination, the large polarity of the surface  is compensated by a surface reconstruction where one subsurface V atom moves from the second layer to the first (surface) layer so that (O$_3$-V) (V-O$_3$-V) (V-O$_3$-V)$\cdots$ actually becomes   (O$_3$-V$_3$-O$_3$) (V-O$_3$-V)$\cdots$, the surface reconstruction for the VO termination is less dramatic.

After atomic relaxation we Wien2Wannier-project \cite{12} the corresponding Wien2K \cite{10} bandstructure and supplement it by a local Kanamori interaction with intra-orbital  $U=\,5.5$ eV, inter-orbital $U'=4.1\,$ eV and Hund's exchange  $J=0.7\,$ eV. The interaction parameters are taken a bit larger than in the literature \cite{Held2001B,Keller2004,Poteryaev2007,tomczak09} first because of the reduced screening at the surface and second to account for the more insulating nature at the experimental 1.1\% Cr-doping, which is too small to take into account in the supercell of our calculations. The resulting Hamiltomian is then solved by DMFT at room temperature, 300 K, using continuous-time quantum Monte Carlo (CTQMC) simulations in the hybridization expansions \cite{14}  (w2dynamic code \cite{15}) and the maximun entropy method \cite{16} for an analytic continuation of the spectra.

{\em DFT+DMFT results.} Fig.\ 4 (lower panels) shows the DFT+DMFT spectra  for VO (left) and  O$_3$  termination (right), resolved for the $a_{1g}$ and $e_g^{\pi}$ orbitals and the different layers. The VO terminated surface is insulating at this interaction strength and temperature. Compared to a 3d$^2$ electronic configuration for all V atoms, the vanadyl termination adds 1 extra  O to each (V-O$_3$-V) unit cell in the layer, i.e., 1 hole per V in the layers. According to our DFT+DMFT results, this hole is however bound to the surface layer where the single V in the surface layer is in a V$^{4+}$ or 3d$^1$ configuration and to the second layer where one of the two V atoms is  V$^{4+}$. The other V atom in the second layer and all other V in the further subsurface layers are in a  V$^{4+}$ or 3d$^2$ configuration. This charge disproportionation explains why, despite of the doping, the surface may remain insulating. (Note,  Fig. 4 show the layer-averaged spectrum, the supplemental material  resolves it for  the two inequivalent sites of the second layer.) If we reduce the  interaction strength, the VO terminated surface becomes metallic. Indeed the VO terminated surface is more metallic than bulk  V$_2$O$_3$, i.e., it stays metallic up to larger values of the Coulomb interaction (see Supplemental material)

The O$_3$ termination  (Fig.\ 4 right) is, on the other hand, already metallic at the same Coulomb interaction and temperature. The most metallic layer is the surface layer; and the width of the central quasiparticle peak shrinks from layer-to-layer. The fourth, central layer  is already close to the bulk result (see Supplemental Material).  The reason why the surface layer is more metallic, despite the reduced hopping, is  the even larger hole-doping due to the O$_3$ termination. The (O$_3$-V$_2$) slab adds to the stoichiometric (V-O$_3$-V) layers in the case of the O$_3$ termination  3 holes per (V-O$_3$-V) unit cell in the layers (or 1.5 holes per V atom). These 1.5 holes are now however distributed to  three surface layers in the DFT+DMFT charge distribution: 1.2 electrons per V (first layer),  1.5  (second layer), 1.9 (third layer), 2.00  (i.e., the bulk value for the fourth layer). That is for the O$_3$ termination Vanadium is neither V$^{4+}$  nor V$^{3+}$ but in between. Consequently, the system is more itinerant;   the surface layers of V$_2$O$_3$ are metallic.

Hence we can  conclude that  excess oxygen makes the (reconstructed) surface more metallic due to hole-doping. The additional surface at the steps indicates an extra surface doping so that V$_2$O$_3$ will be even more metallic at such edges. This is akin to our O$_3$ termination vs.\  VO termination.  As we have seen, this larger amount of doping has not a very big effect, but it shifts the critical Coulomb interaction for the Mott-Hubbard transition somewhat  to the left of the phase diagram (inset Fig. \ref{spectra}). That means the transition temperature increases as we found experimentally.  It also explains  our experiment, Fig. \ref{defects}; the cleave step is more metallic because of an accumulation of excess oxygen at the corner.

\begin{figure}
\includegraphics[angle=0,width=1\linewidth,clip=true]{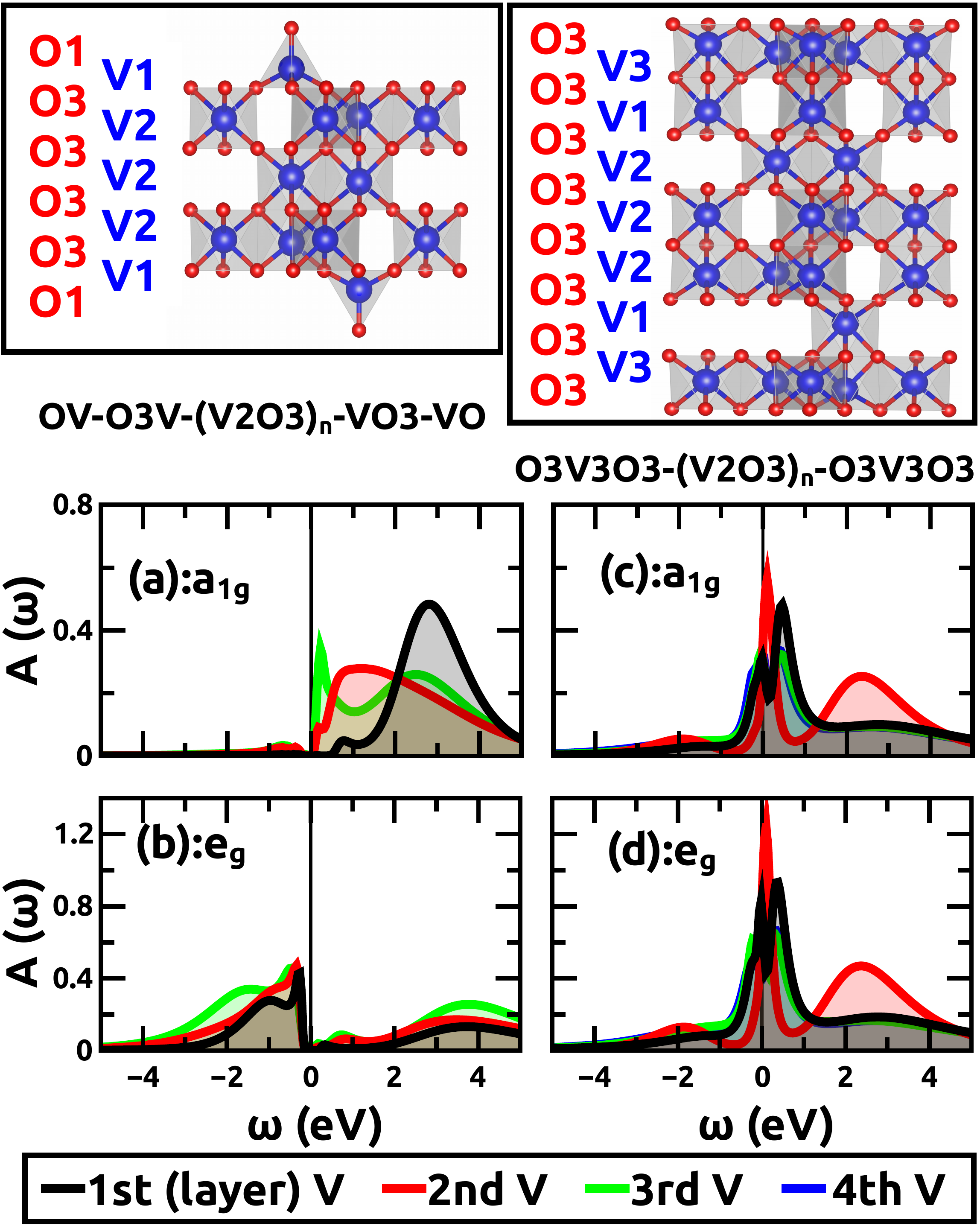}
\caption{{\bf Top:} VO termination (left) and Supercell  O$_3$ termination (right). Below the figures, we indicate how the supercell is made up from the stoichiometric (V-O$_3$-V) supercell plus the surface termination. For the O$_3$ termination, one V moves from the second to the first layer because of the surface reconstruction.
{\bf Bottom:} Layer- and orbital-resolved spectral function $A(\omega)$ {\bf  (a)-(b)} for the VO  termination, 
{\bf (c)-(d)} for the (O$_3$) termination.
The more excess  oxygen at the surface, the more metallic are the surface layers.} 
\label{dmft} 
\end{figure}

{\em Conclusions.} We observed experimentally that at the PI/PM Mott transition in (V$_{1-x}$Cr$_x$)$_2$O$_3$ metallic domains appear at higher temperatures than the bulk transition. Their evolution is determined by the surface crystallographic direction and along the cleaving steps. Our DFT+DMFT theoretical calculations show that a surface reconstruction with an excess of oxygen favors the formation of a metallic phase. Against common wisdom, sufaces can hence be more metallic than the bulk, and surface steps even more so. This effect observed here for (V$_{1-x}$Cr$_x$)$_2$O$_3$ can be of general interest for surfaces of strongly correlated oxides, oxide heterostructures, and nanostructure. It is also relevant when comparing surface and bulk sensitive experiments.

{\em Acknowledgments.} LS and KH acknowledge financial support by European Research Council under the European Union's Seventh
Framework Program (FP/2007-2013)/ERC through grant agreement n.\ 306447 and 
the Doctoral School W1243   Solids4Fun  by the Austrian Science Fund (FWF). Calculations have been done on the Vienna Scientific Cluster~(VSC). GL, MH, EP, and MM acknowledge financial support by the EU/FP7 under the contract Go Fast (Grant No. 280555) and the  Labex PALM. The SPEM experiments have received funding from the European Community’s Seventh Framework Program (FP7, grant No 312284).

\bibliographystyle{apsrev}

\end{document}